\documentstyle[11pt,newpasp,twoside,psfig,amstex]{article}
\begin{document}
\title{Variations in the Bivariate Brightness Distribution with different 
galaxy types.}
\author{Nicholas Cross}
\affil{Department of Physics and Astronomy, Johns Hopkins University, 3700, 
San Martin Dr, Baltimore, MD 21218, USA.}
\author{Simon Driver}
\affil{Research School of Astronomy \& Astrophysics, Mount Stromlo Observatory,
Cotter Road, Weston ACT 2611, Australia.}
\author{David Lemon, Jochen Liske}
\affil{School of Physics \& Astronomy, The North Haugh, St Andrews, Fife,
KY16 9SS, United Kingdom.}

\begin{abstract}
We present Bivariate Brightness Distributions (BBDs) for four spectral types 
discriminated by the 2dFGRS. We discuss the photometry and completeness of 
the 2dFGRS using a deep, wide-field CCD imaging survey. We find that there is 
a strong luminosity-surface brightness correlation amongst galaxies with 
medium to strong emission features, with gradient 
$\beta_{\mu}=0.25\pm0.05$ and width $\sigma_{\mu}=0.56\pm0.01$. Strong 
absorption line galaxies, show a bimodal distribution, with no correlation 
between luminosity and surface brightness. 
\end{abstract}

\vspace{-10mm}
\section{Introduction}

The BBD is the number density of galaxies as a function of both absolute 
magnitude, $M$, and effective surface brightness, $\mu_e$. This provides a 
useful framework for removing surface brightness selection effects from the
measurement of the space density of galaxies (Cross et al. 2001) and also 
provides useful constraints for testing models of galaxy formation and 
evolution (de Jong \& Lacey 2000).

Large redshift surveys such as the ``Two-degree Field Galaxy Redshift Survey''
(2dFGRS, Colless et al. 2001) and the ``Sloan Digital Sky Survey (SDSS, 
York et al. 2000) provide the quantity of data necessary to select the
unbiased, statistically significant samples needed to measure the BBDs of
different galaxy types. 

\vspace{-5mm}
\section{The Data}

The 2dFGRS is a $2000\deg^2$ redshift survey, containing redshifts 
for $\sim 220,000$ galaxies, with an apparent magnitude limit $b_j=19.45$ 
(see Colless et al. 2001). The input catalogue for the 2dFGRS is based on the 
Automated Plate Machine (APM) catalogue (Maddox et al. 1990), with some recent
recalibration using new CCD data. 
 
Madgwick et al. (2002) used the 2dFGRS to identify four spectral classes 
based on the emission/absorption line strength ($\eta$) of galaxies.
$\eta$-type 1 galaxies have strong absorption lines and $\eta$-type 4 
galaxies have strong emission lines.

The Millennium Galaxy Catalogue (MGC, Liske et al. 2002) is a 
$\sim36$deg$^{2}$ deep CCD imaging survey of an equatorial strip between 
$10^h$ and $15^h$. The MGC images were taken through the KPNO-B filter on the 
Isaac Newton Telescope. The detection isophote $\mu_{lim}=26.0$ mag 
arcsec$^{-2}$ and the mean seeing is $1.3\arcsec$. Star-galaxy separation has 
been undertaken by eye for all $B_{MGC}<20$ objects. For each 
galaxy we measured half light radii ($r_e$) along the major axis of the 
ellipse that contains half of the flux of the galaxy. We use this to 
calculate the effective surface brightness, $\mu_e=B_{MGC}+2.5\log_{10}(2
\pi\,r_e^2)$, assuming a circular aperture, to correct for inclination 
effects, see Cross et al. (2002).

\vspace{-5mm}
\section{Photometry and Completeness of the 2dFGRS}

Using a mean colour for galaxies $(\overline{B-V})=1.0$, we find that the 
filter conversion between the APM and MGC is $B_{MGC}-b_j=0.1353(B-V)=0.1353$.
$\Delta\,m(MGC-2dFGRS)=-0.078\pm0.157$.

\begin{figure}
\psfig{file=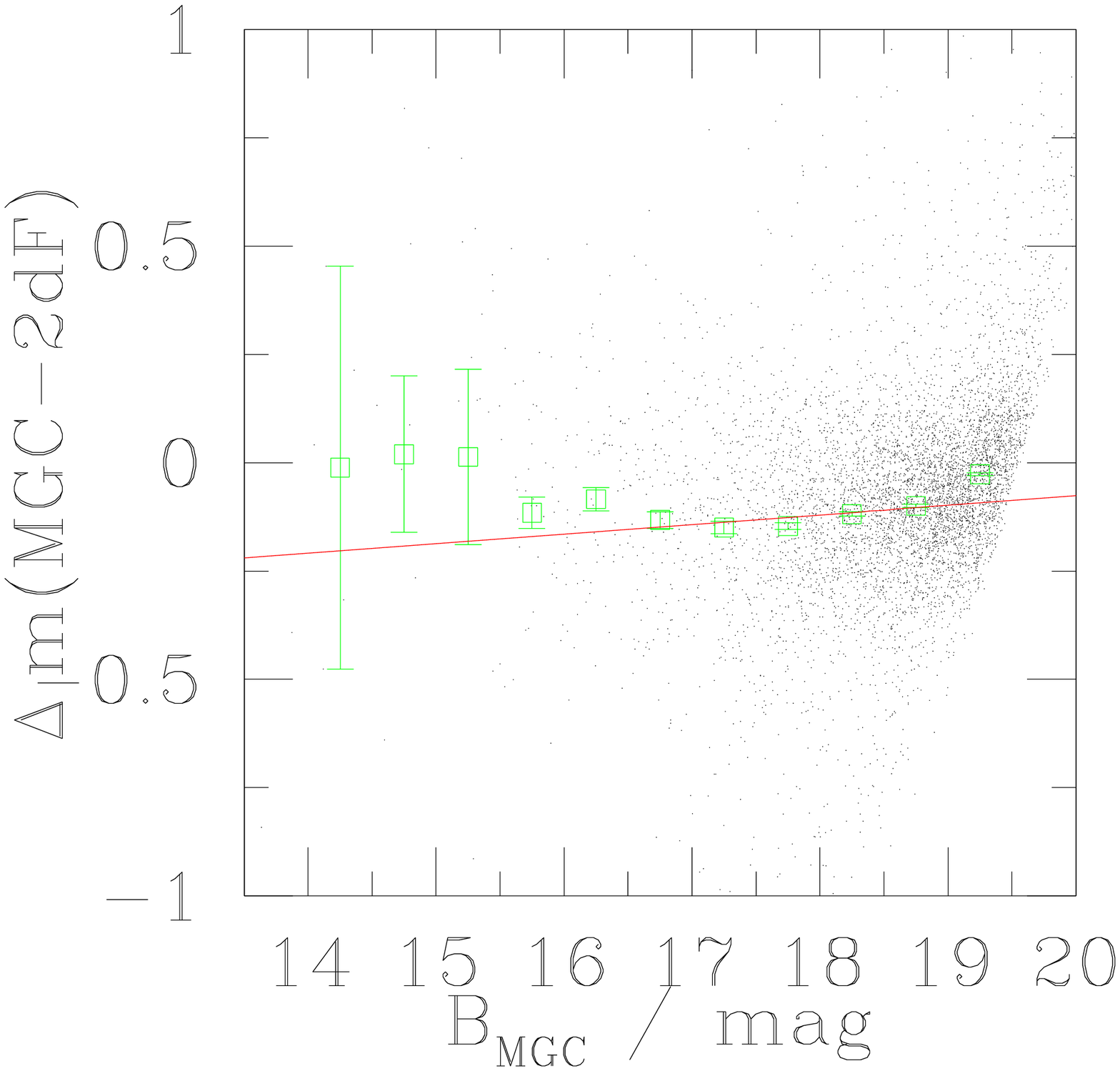,width=30mm,height=30mm}
\vspace{-30mm}
\hspace{30mm}
\psfig{file=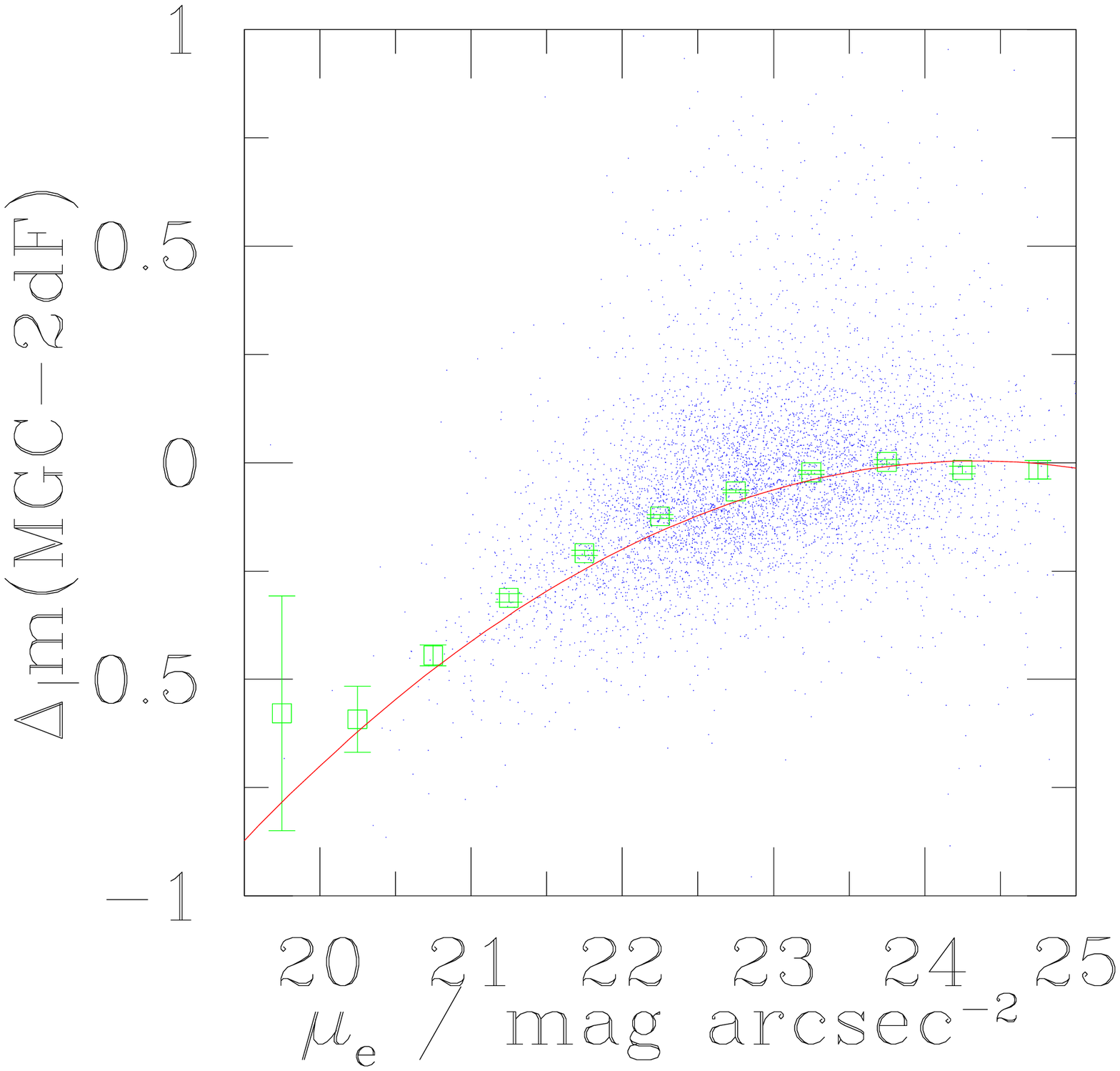,width=30mm,height=30mm}
\psfig{file=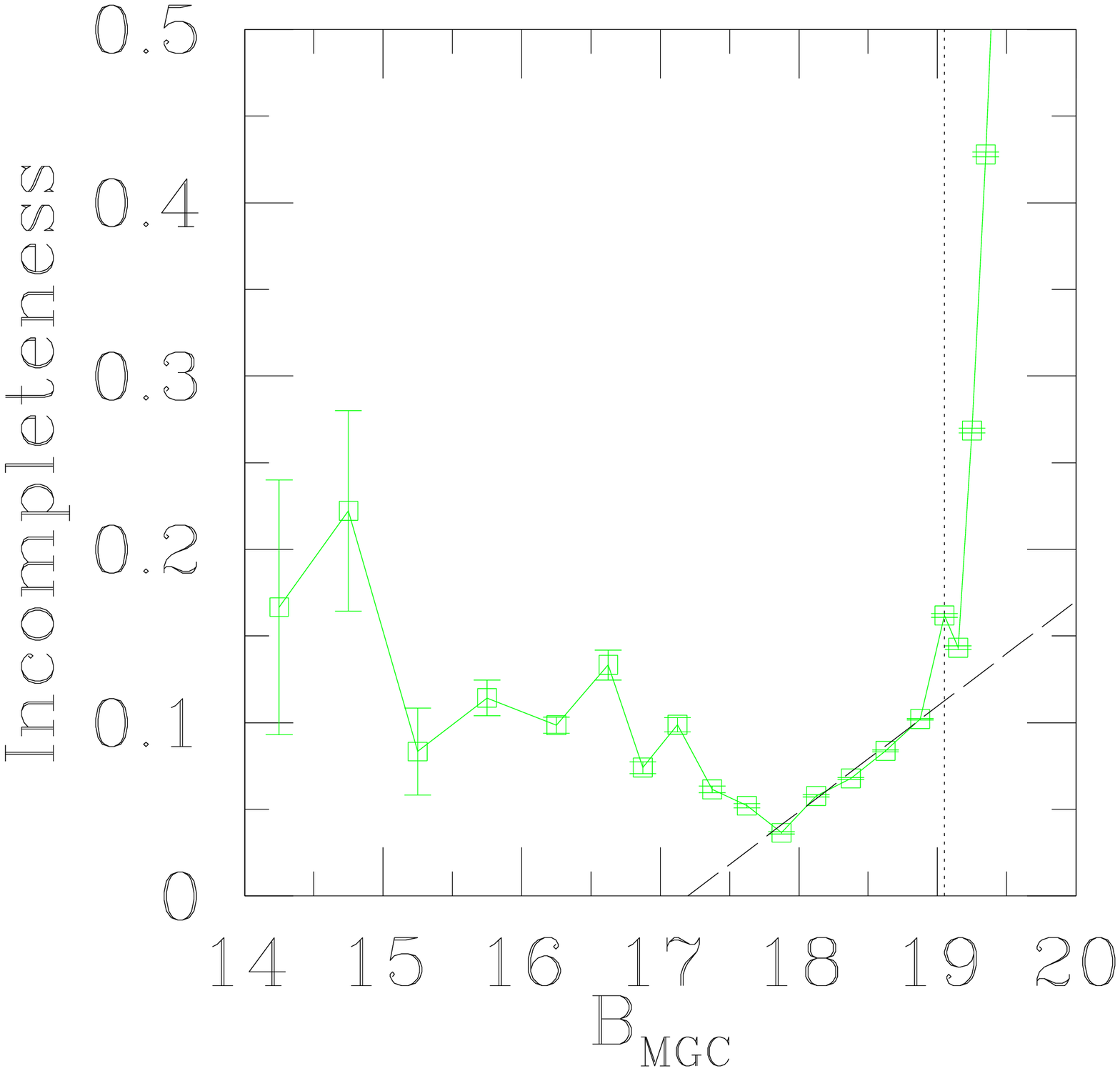,width=30mm,height=30mm}
\psfig{file=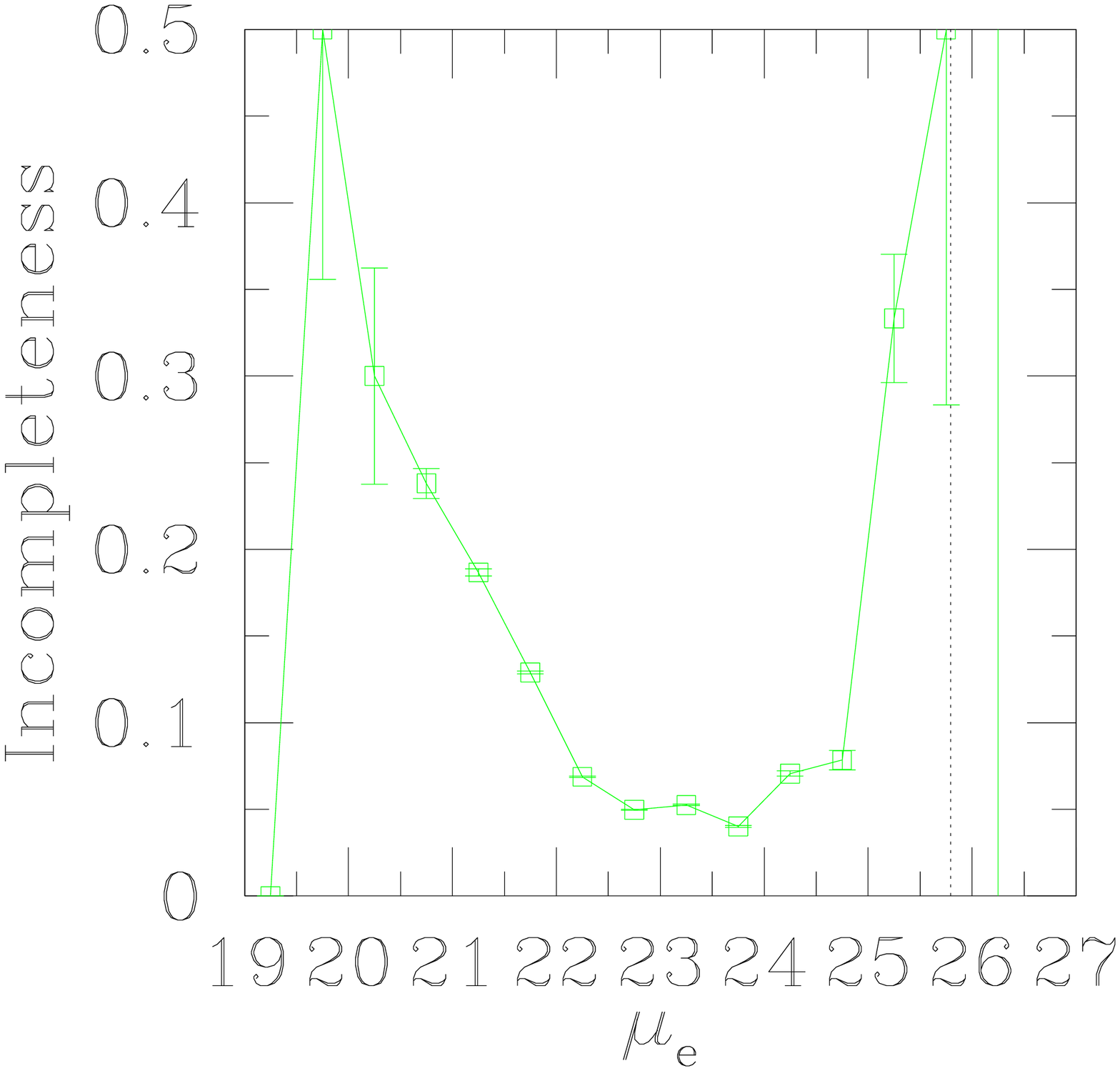,width=30mm,height=30mm}
\caption{The photometry and completeness of the 2dFGRS. The left-hand plots 
show the scale error with magnitude, and surface brightness. The rightmost 
plots show the incompleteness as a function of magnitude and surface 
brightness respectively.
\label{fig:2dF_MGC}}
\end{figure}

The mean scale error in magnitude is small and linear (see 
Fig~\ref{fig:2dF_MGC}). It is fit by the equation 
$\Delta\,m=a+b(B_{MGC}-19.45)$, where $a=0.0467$ and $b=0.0221$. However, 
there is a significant scale error as a function of effective surface 
brightness (see Fig~\ref{fig:2dF_MGC}). $\Delta\,m$ is constant
for $22<\mu_e<25$ but decreases non-linearly with $\mu_e$ for $\mu_e<22$ mag
arcsec$^{-2}$. The most likely explanation is non-linearities or saturation 
of high surface brightness galaxies in the APM. A comparison between the MGC 
and the Sloan Digital Sky Survey Early Data 
Release (SDSS-EDR) shows that $\Delta\,m(MGC-SDSS)=-0.009\pm0.100$ with no
significant scale error. 

We select the sample of MGC galaxies which should have 2dFGRS matches to 
measure the photometric completeness, see Cross et al. (2002). The variation 
of incompleteness with magnitude and with surface brightness is shown in 
Fig.~\ref{fig:2dF_MGC}. The incompleteness is worst at the bright and 
faint ends, reaching a minimum of $\sim5\%$ in the range $17.25<B_{MGC}<18.0$.
It increases linearly with magnitude in the range $18.0<B_{MGC}<19.0$. For 
$B_{MGC}>19.0$, the magnitude limit of the 2dFGRS affects the measurement of 
completeness. For 
$22.25<\mu_e<24.25$ the incompleteness is fairly constant $\sim5\%$. The 
incompleteness of low surface brightness galaxies increases rapidly beyond
$\mu_e=24.25$ and no 2dFGRS galaxies are seen with $\mu_e>25.75$, as expected 
with an isophotal limit $\mu_{lim}=24.67$. At the high surface brightness 
end, the incompleteness rises steadily, since small angular size galaxies were
mistaken for stars in the APM.

\vspace{-5mm}
\section{Bivariate Brightness Distributions}

Since the 2dFGRS suffers from a non-linearity with surface brightness and 
$8\%$ photometric incompleteness to $B_{MGC}>19.0$, we use the MGC objects 
defined as galaxies to select the BBD sample. The selection limits used are:
the isophotal limit $\mu_{lim,B}=26.0$ mag arcsec$^{-2}$, the minimum half 
light radius $r_e=0.793''$, defined by the average star-galaxy separation, 
and the maximum half-light radius $r_e=15.8''$, limited by the sky
subtraction. The other limits, $15.0<B_{MGC}<19.0, 21.0<\mu_e<24.5$ are 
determined by the redshift completeness. We select a subsample, with $>70\%$ 
redshift completeness at all values of $B_{MGC}$ and $\mu_e$. These limits are
discussed in Cross (2002).

\begin{figure}
\psfig{file=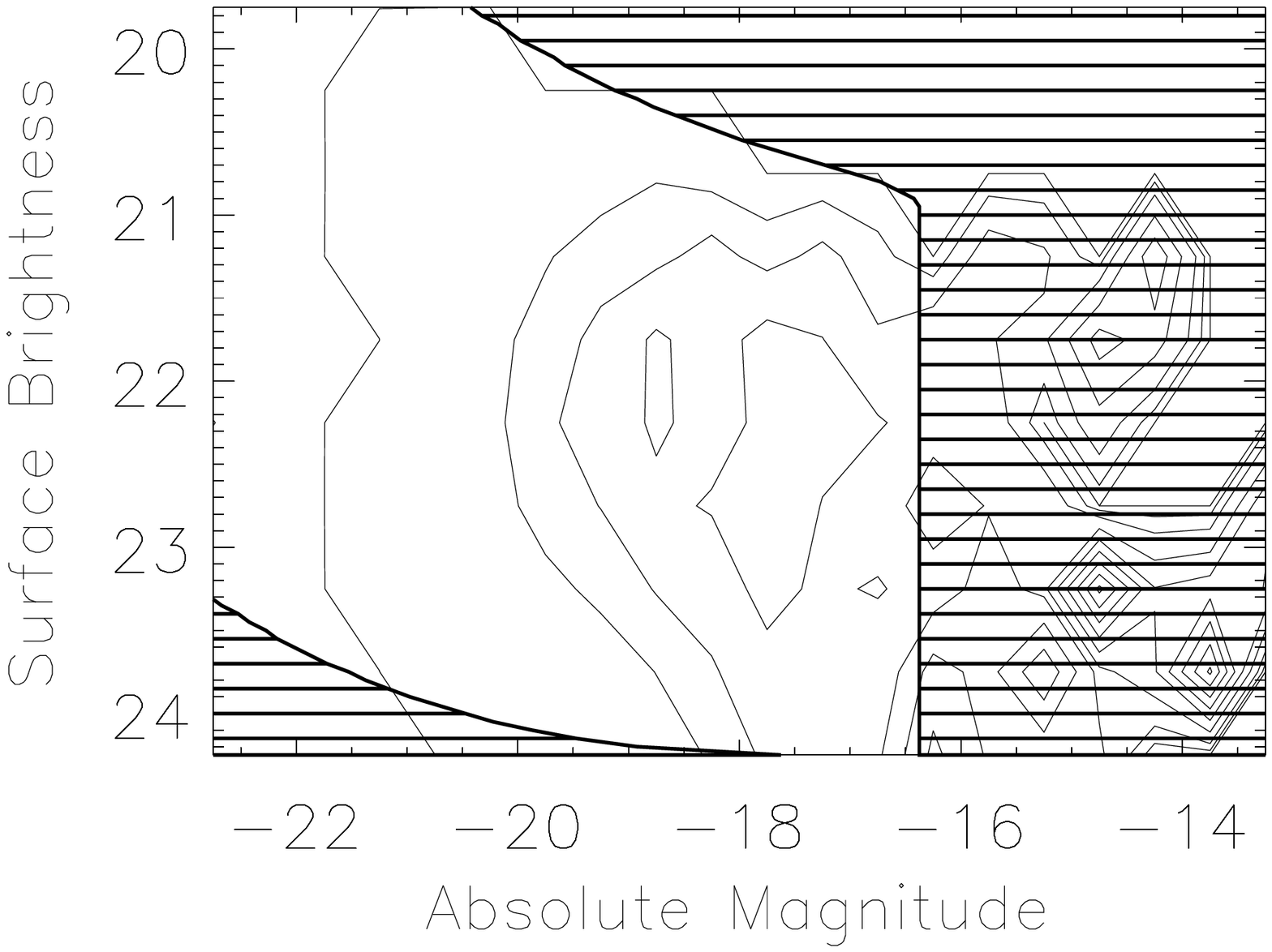,width=25mm,height=25mm}
\vspace{-25.0mm}
\hspace{25.0mm}
\psfig{file=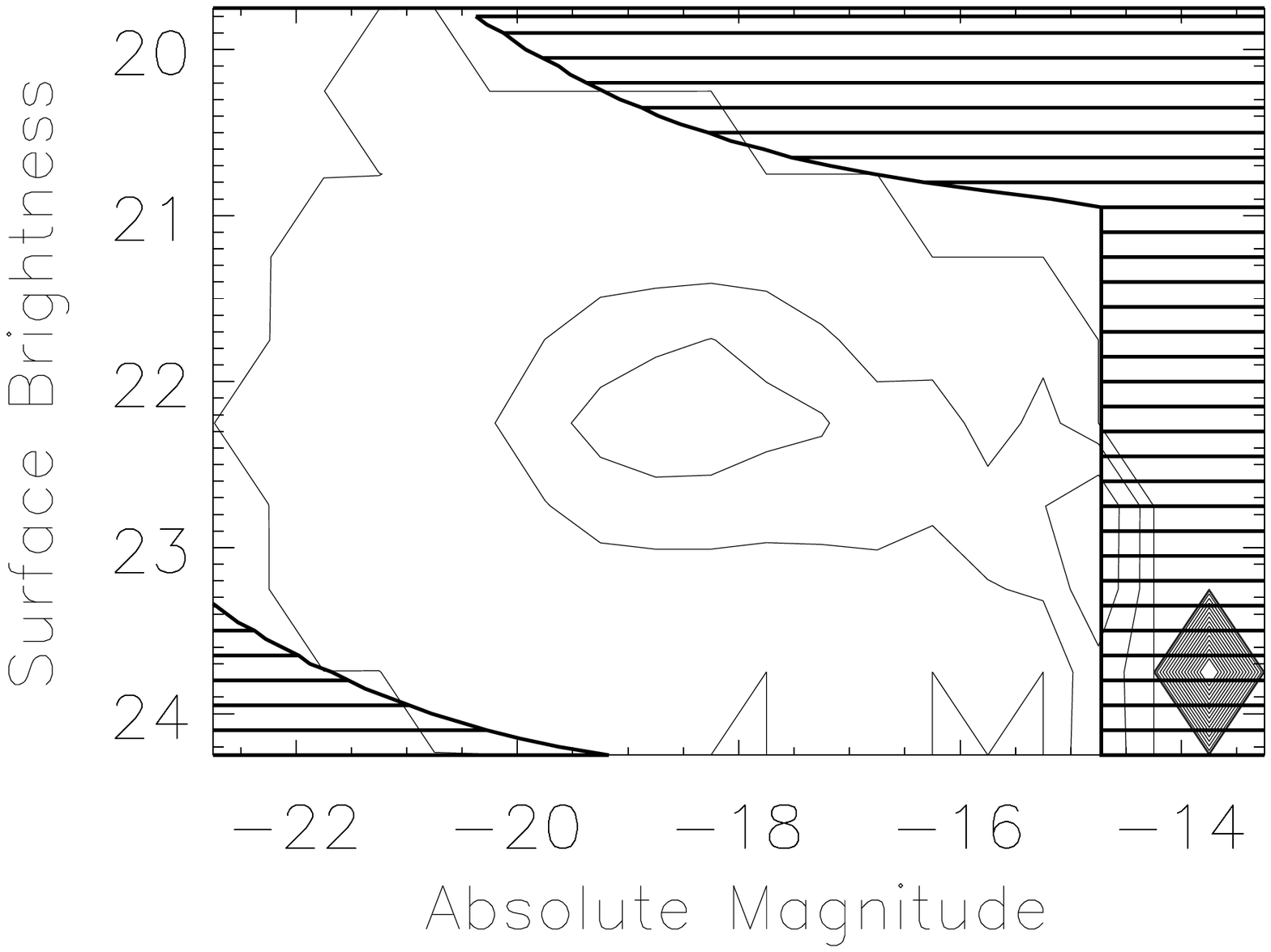,height=25.0mm,width=25.0mm}
\psfig{file=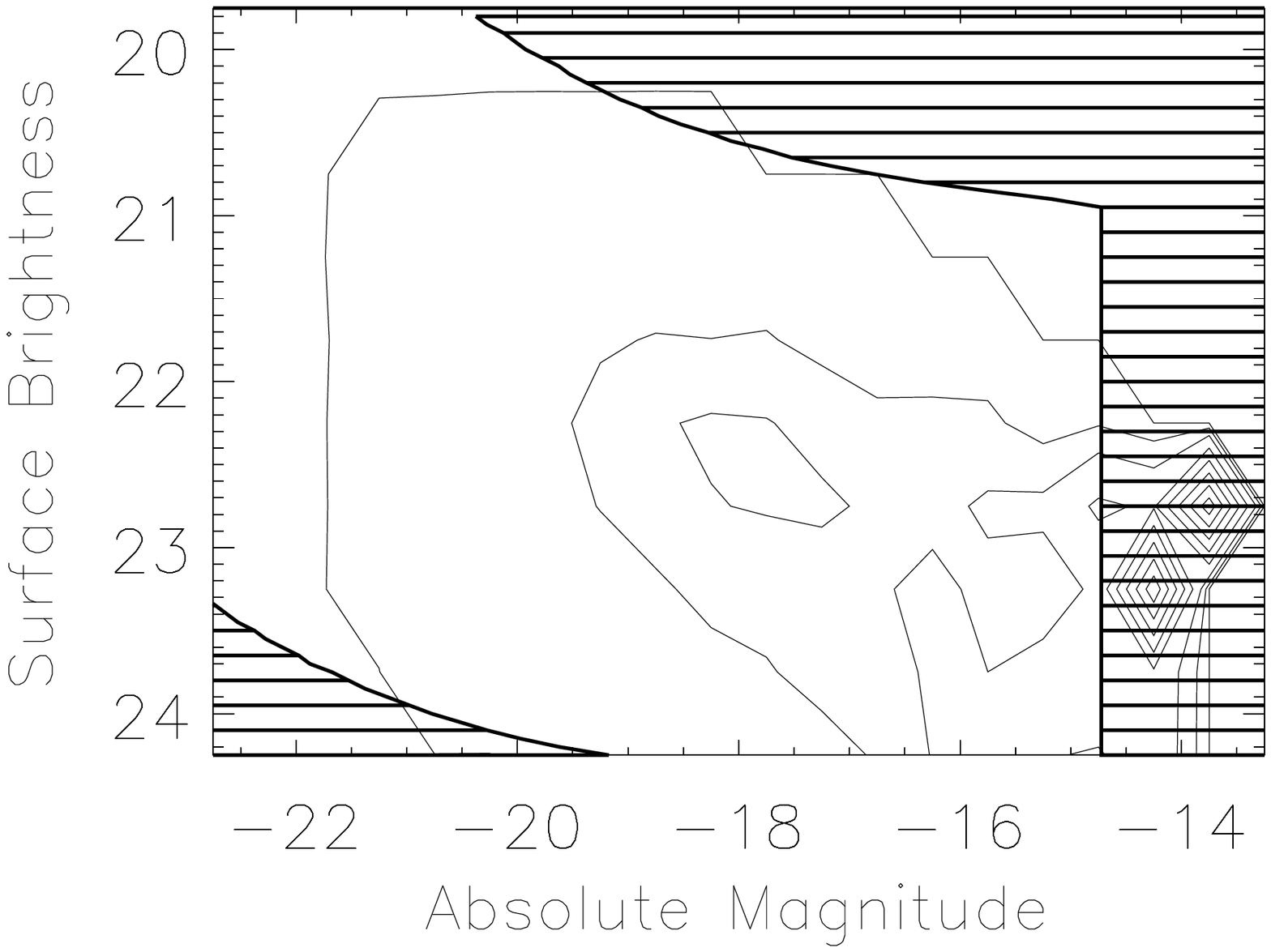,height=25.0mm,width=25.0mm}
\psfig{file=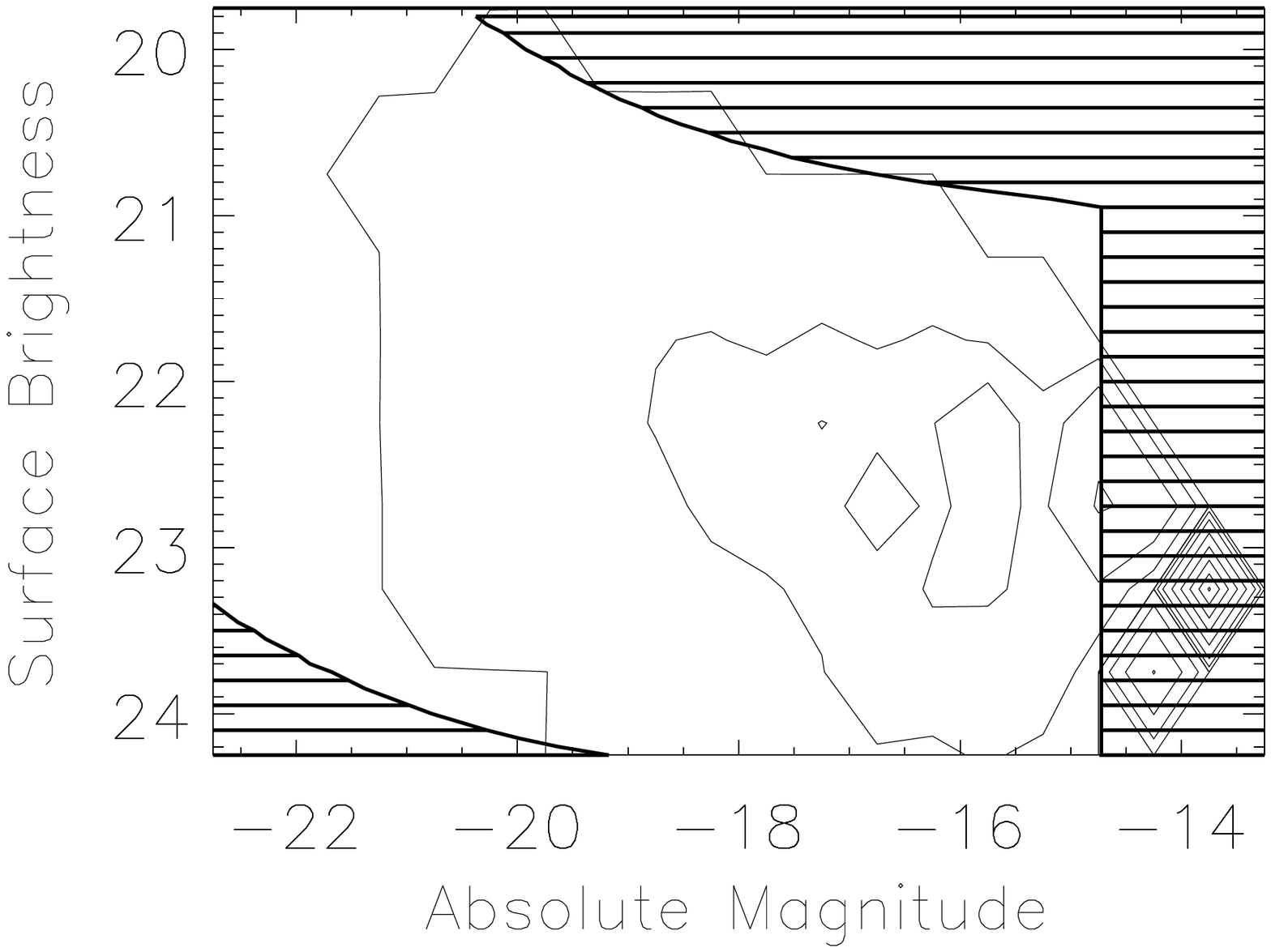,height=25.0mm,width=25.0mm}
\psfig{file=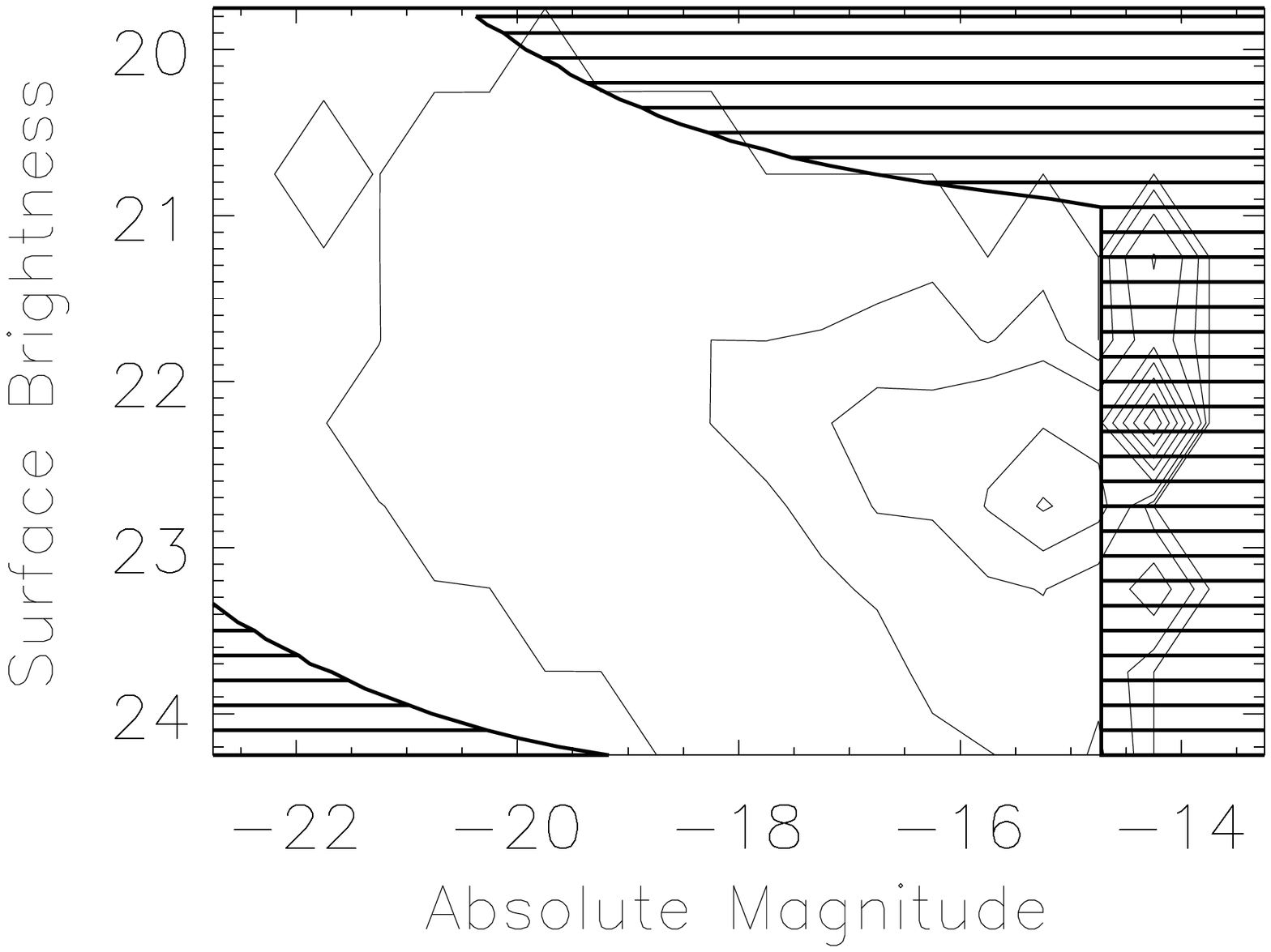,height=25.0mm,width=25.0mm}
\caption{The Bivariate Brightness Distribution for MGC/2dFGRS galaxies. The 
first plot shows the BBD produced for the whole galaxy distribution. The
other 4 plots from left to right show the BBDs for $\eta$-type 1, 2, 3 and 4
galaxies respectively. The contour lines for the BBDs are set at 
$1.0\times 10^{-7}$, $1.0\times 10^{-3}$, $2.5\times 10^{-3}$, and then 
increasing in steps of $2.5\times 10^{-3}$ galaxies Mpc$^{-3}$ bin$^{-1}$.
The shaded region marks the region where the volume, $V<5000$ Mpc$^3$, is too
small to accurately measure the space density.}
\label{fig:BBD}
\end{figure}

We calculate the overall BBD using a maximum likelihood method outlined in 
Cross (2002). It is based on the step-wise method (Efstathiou, Ellis \& 
Peterson 1988), but also includes Visibility Theory (Phillipps, Davies \& 
Disney 1990). The overall BBD is shown in the first plot of Fig~\ref{fig:BBD}.

\begin{equation}
%\begin{split}
\phi(M,\mu_e)=\frac{0.4\ln(10)}{\sqrt{2\pi}\,\sigma_{\mu}}\phi_*\,
10^{0.4(M^*-M)(\alpha+1)}\,e^{-10^{0.4(M^*-M)}}e^{\left[-\frac{1}{2}\left(\frac{\mu_e-\mu_e^*-\beta_{\mu}\,(M-M_*)}{\sigma_{\mu}}\right)^2\right]}  
\label{eq:func}
%\end{split}
\end{equation}

We fit the Cho{\l}oniewski (1985) function, Eqn~\ref{eq:func}, to this 
distribution. The best fit gives $M^*=-19.24$, $\phi^*=2.47\times10^{-2}$,
$\alpha=-0.796$, $\beta_{\mu}=0.180$, $\mu_e^*=22.06$ and 
$\sigma_{\mu}=0.765$ with $\chi^2=220$ for $\nu=101$ degrees of freedom.

The MGC database is too small to produce statistically significant samples to
measure the BBD for each spectral type. Instead, we use the MGC to remove the 
non-linearities in the 2dFGRS magnitudes and to correct for photometric
incompleteness and stellar contamination, see Cross (2002). This gives a ten 
times larger dataset, but with larger random errors in the photometry.

The BBDs produced for each type are also displayed in Fig~\ref{fig:BBD}. 
We find that $\eta$-type 1 galaxies have a bimodal distribution - a 
tightly bounded distribution, centred on $M=-18.5,\mu_e=22.2$ with a high 
degree of symmetry and no significant luminosity surface brightness 
correlation, with a second fainter distribution, which continues upto the 
selection limits. 

In contrast types 2-4 show a strong luminosity-surface brightness correlation
($\beta_{\mu}=0.25\pm0.05$, $\sigma_{\mu}=0.56\pm0.01$), which does not vary 
significantly between these types. The proportion of faint galaxies 
increases with $\eta$-type: $\eta$-type 2 galaxies have a flat faint 
end slope ($\alpha=-0.8$) and $\eta$-type 4 galaxies have a steep faint 
end slope ($\alpha=-1.5$).

\vspace{-5mm}
\section{Conclusions}

We have used a deep, wide field CCD survey, the MGC, to show that the 2dFGRS 
contains a scale error with surface-brightness, due to non-linearities in the 
photographic plates. Furthermore, the 2dFGRS has a mean photometric 
incompleteness of $8\%$, for $B_{MGC}<19$, with many low surface brightness and
low angular resolution galaxies excluded.

The MGC can be used to define a dataset with well understood selection limits
that contains CCD photometric data and 2dFGRS spectroscopic data. We have
used this dataset to calculate the bivariate brightness distribution (BBD) for
the overall galaxy distribution and for four spectral types.

We show that strongly absorbing galaxies (ellipticals/lenticulars) form a 
bimodal population. The bright, high surface brightness population does not 
show a strong correlation between magnitude
and surface brightness. $\eta$-type 2-4 galaxies (spirals/irregulars) all 
show the same strong correlation between magnitude and surface brightness. 
The proportion of dwarf galaxies increases with increasing emission line 
strength.

\end{document}